\documentclass{emulateapj}

\usepackage{apjfonts}

\catcode`\@=11
\newcommand{\gapprox}{\mathrel{\mathpalette\@versim>}}
\newcommand{\lapprox}{\mathrel{\mathpalette\@versim<}}
\newcommand{\propapprox}{\mathrel{\mathpalette\@versim\propto}}
\newcommand{\@versim}[2]
  {\lower3.1truept\vbox{\baselineskip0pt\lineskip0.5truept
\ialign{$\m@th#1\hfil##\hfil$\crcr#2\crcr\sim\crcr}}}
\catcode`\@=12

\shorttitle{DISCOVERY OF PULSATIONS AND A POSSIBLE SPECTRAL FEATURE IN THE X-RAY EMISSION FROM RRAT J1819--1458}
\shortauthors{MCLAUGHLIN ET AL.}

\begin{document}

\title{Discovery of Pulsations and a Possible Spectral Feature in the \\ X-ray Emission from Rotating Radio Transient  
J1819--1458}

\author{M.~A.~McLaughlin,\altaffilmark{1,2} N.~Rea,\altaffilmark{3,4}
B.~M.~Gaensler,\altaffilmark{4} S.~Chatterjee,\altaffilmark{4} 
F.~Camilo,\altaffilmark{5}\\ M.~Kramer,\altaffilmark{6}  D.~R.~Lorimer,\altaffilmark{1,2}
A.~G.~Lyne,\altaffilmark{6} G.~L.~Israel,\altaffilmark{7} \& A.~Possenti\altaffilmark{8}} 

\altaffiltext{1}{Department of Physics, West Virginia University,
Morgantown, WV 26501.}
\altaffiltext{2}{National Radio Astronomy Observatory, Green Bank, WV 24944.}
\altaffiltext{3}{SRON -- Netherlands Institute for Space Research,
   Sorbonnelaan, 2, 3584 CA, Utrecht, The Netherlands.}
\altaffiltext{4}{School of Physics, The University of Sydney, NSW 2006, Australia }
\altaffiltext{5}{Columbia Astrophysics Laboratory, Columbia University,
New York, NY 10027.}
\altaffiltext{6}{Jodrell Bank Observatory, University of Manchester,
   Macclesfield, Cheshire SK11 9DL, UK.}
\altaffiltext{7}{INAF -- Osservatorio Astronomico di Roma,
 I-00040 Monteporzio Catone, Italy.}
\altaffiltext{8}{INAF -- Osservatorio Astronomico di Cagliari,
Loc. Poggio dei Pini, 09012 Capoterra, Italy.}

\begin{abstract}

PSR~J1819--1458 is a rotating radio transient (RRAT) source with
  an inferred surface dipole magnetic field strength of
  $5\times10^{13}$\,G and a 4.26-s spin period. We present 
 {\it XMM-Newton} observations of the X-ray counterpart of this source, CXOU~J181939.1--145804,
in which we identify pulsations and a possible spectral
  feature.
  The X-ray pulsations are
  at the period predicted by the radio ephemeris, providing
  an unambiguous identification with the radio source and confirmation
  of its neutron star nature. The X-ray pulse has a 0.3--5~keV pulsed
  fraction of 34\% and is aligned with the expected phase of the radio pulse.  The X-ray
  spectrum is fit well by an absorbed blackbody with $kT = 0.14$~keV
  with the addition of an absorption feature at 1\,keV, with total
  absorbed flux of $1.5\times10^{-13}$~ergs~cm$^{-2}$~s$^{-1}$
  (0.3--5~keV).  This absorption feature is well modeled by a Gaussian
  or resonant cyclotron scattering model, but its significance is
  dependent on the choice of continuum model.  We find no
  evidence for any X-ray bursts or aperiodic variability on timescales
  of 6~ms to the duration of the observation and can place the most
  stringent limit to date of $\le
  3\times10^{-9}$~ergs~cm$^{-2}$~s$^{-1}$ on the absorbed 0.3--5~keV
  flux of any bursts.

\end{abstract}


\section{Introduction}
\label{intro}

In 2006 February, a new class of neutron stars, the ``Rotating RAdio
Transients'' (RRATs) was reported (McLaughlin et al. 2006). These 11
objects, characterized by repeated dispersed radio bursts, have
periods ranging from 0.7 to 7 seconds and are located in the Galactic
plane at 2 -- 7 kpc distances.  Their periods are longer than those of
most normal radio pulsars and similar to those of the populations
of X-ray dim isolated neutron stars (XDINS; see Haberl~2007 for a
review) and magnetars (see Woods \& Thompson~2006 for a review).  For
the three sources with the highest pulse detection rates, period
derivatives, $\dot P$, have been determined.  If the measured $\dot P$
values are interpreted as due to magnetic dipole spin-down, they imply
characteristic ages and magnetic field strengths in the general range
of the normal pulsar population.

There have been several suggestions put forward on the nature of this
new class of neutron star. One obvious suggestion is that the RRATs
are related to pulsars which emit ``giant pulses'' (e.g. Knight et al. 2006).
However, the RRATs
with measured period derivatives do not appear to have high values of
either magnetic field strength at
the light cylinder or spin-down luminosity, both suggested as predictors of
giant-pulse activity (Cognard et al. 1996; Knight et al. 2006).
 Zhang et al. (2006) suggest that
the RRATs may be neutron stars near the radio ``death line'' or may be
related to ``nulling'' radio pulsars. However, the period derivatives
measured for three RRATs do not place them near canonical pulsar
``death lines'' (e.g. Chen \& Ruderman 1993) and, unlike most nulling
pulsars (e.g. Wang et al. 2007), we typically do not see more than one
pulse from the RRATs in succession.  Another intriguing possibility is
that the sporadicity of the RRATs is due to the presence of a
circumstellar asteroid belt (Cordes \& Shannon 2006; Li 2006) or a
radiation belt such as seen in planetary magnetospheres (Luo \&
Melrose 2007). Or, perhaps, they are transient X-ray magnetars, a
particularly relevant suggestion given the recent detection by Camilo
et al. (2006) of transient radio pulsations from the anomalous X-ray
pulsar XTE~J1810--197. A final possibility is that they are similar
objects to PSR~B0656+14, one of three middle-aged pulsars (i.e. ``The
Three Musketeers''; Becker \& Truemper 1997) from which pulsed
high-energy emission has been detected (e.g. DeLuca et al. 2005).
Weltevrede et al. (2006) convincingly show that if PSR~B0656+14 were
more distant its emission properties would appear similar to those of
the RRATs.  Determining the reason for the unusual emission of the
RRATs is important as population analyses show that their population
may be up to several times greater than that of the normal radio
pulsars (McLaughlin et al. 2006).  Popov et al. (2006) show that the
inferred birthrate of RRATs is consistent with that of XDINS but not
with magnetars.

While the radio emission properties of J1819--1458 are quite different
from those of ``normal'' pulsars, it appears to be a rotating neutron
star from which we detect radio pulses, and we henceforth give it the
prefix ``PSR''.  PSR~J1819--1458 shows the brightest and most frequent
radio bursts of any of the RRAT sources. It has a 4.26-s period,
relatively high inferred characteristic surface dipole magnetic field
strength of $5\times10^{13}$~G, characteristic age 
of 117~kyr, and spin-down luminosity of
$3\times10^{32}$~ergs~s$^{-1}$. The distance inferred from its
dispersion measure (DM) of 196$\pm3$~pc~cm$^{-3}$ is 3.6~kpc (Cordes
\& Lazio~2002), with considerable (at least 25\%) uncertainty.  PSR~J1819--1458 is
characterized by radio bursts at 1.4~GHz of average duration 3~ms,
with bursts arriving randomly with a mean rate of one every $\sim$~3~minutes.  X-ray emission was
detected from this source in a serendipitous 30-ks {\it Chandra}
ACIS-I observation toward the Galactic supernova remnant G15.9+0.2
(Reynolds et al. 2006)\footnote{As discussed by Reynolds et
  al. (2006), an association between the remnant and PSR~J1819--1458
  is extremely unlikely.}.  They found that the spectrum was well
described by an absorbed blackbody with neutral hydrogen column
density $N_H = 7^{+7}_{-4} \times 10^{21}$ cm$^{-2}$ and temperature
$kT=0.12 \pm 0.04$\,keV, with an absorbed flux of $\sim1 \times
10^{-13}$\,ergs\,cm$^{-2}$\,s$^{-1}$ between 0.3 and 5 keV. These
properties are consistent with emission from a cooling neutron star of
age $10^4-10^5$~years, broadly consistent with the characteristic age
of PSR~J1819--1458. No evidence for bursts or variability was found.
The time resolution of these {\it Chandra} observations was not
sufficient to allow a robust search for X-ray pulsations.

We were awarded 43~ks of {\it XMM-Newton} time to further characterize
the spectrum and search for pulsations. We report here on the results
of these observations, in particular the detection of X-ray pulsations and a
possible feature in the X-ray spectrum. This is the first
detection of X-ray pulsations from any of the RRAT sources. In
Section~2 we describe the observations and both the timing and
spectral analyses. In Section~3 we discuss possible interpretations of
our results, and present our conclusions in Section~4.

\section{Observations and Analysis}
\label{obs}

The {\it XMM-Newton} observations were performed on 2006 5 April, with
a total observation time of 43~ks. The European Photon Imaging Camera
(EPIC) PN and MOS instruments were operated with medium filters and in
Small Window mode, providing a time resolution of 6 and 300~ms and
effective livetimes of 71\% and 97.5\%, respectively.  The data were
reduced using the {\it XMM-Newton} Science Analysis System (SAS
version 7.0.0) and the most recent calibration files. Data from both
the PN and MOS instruments were used for the timing and spectral
analyses.  We also also analyzed the Reflection Grating Spectrometer data but
our target was too faint to be reliably detected.

We filtered the observation for background flares, resulting in
effective on-source exposure times of 32.5 and 37.8~ks (23.0 and
36.9~ks including deadtime) for the PN and MOS instruments,
respectively.  We detect a point source with J2000 coordinates: right ascension $\alpha
= 18^{\rm h}19^{\rm m} 34^{\rm s}$ and declination $\delta = -14^\circ 58' 03''$ 
(4$^{\prime\prime}$ error in each coordinate). This is consistent with
the radio-timing-derived position
and with the more accurate position for the X-ray counterpart
published in Reynolds et al. (2006). There is no evidence for extended
emission, with the source brightness falling off as expected given the
{\it XMM-Newton} point spread function.  

For both PN and MOS data, we extracted the source photons within a
20$^{\prime\prime}$ circular radius centered on the source position,
which ensured extraction of more than 90\% of the source counts. Following
standard practice, 
the size of the extraction region has not been corrected for in
our final quoted fluxes and luminosities. 
The background counts were extracted from four 20$^{\prime\prime}$
circular regions centered on off-source positions (different for the
PN and MOS instruments) in the same central CCD.

\subsection{Timing Analysis}

For the timing analysis, we used all photons from PN and MOS1 and MOS2 
instruments with PATTERN $\le$ 12 (i.e. allowing for single, double, triple and
quadruple events\footnote{http://xmm.vilspa.esa.es/external/xmm\_user\_support/documentation/uhb}).
 This resulted in $2200\pm58$ and
$945\pm15$ source and background PN counts and $1380\pm41$ and $230\pm8$
source and background MOS1 plus MOS2 counts (normalized for a 20$^{\prime\prime}$ circular extraction
region).
Although we can measure
a period of 4.26~s from the arrival times of the radio bursts, the
periodicity is not detectable in a Fast Fourier Transform of the radio
time series due to the sporadic nature of the source.
We therefore performed a periodicity search of all PN and
MOS photons (i.e. with energies 0.2--15~keV) as a confirmation of the method through which the radio
period was derived (see McLaughlin et al. 2006).  The arrival times
were converted to the solar system barycenter and the $Z_1^2$ test
(Buccheri et al. 1983) was applied. The most significant signal
is detected
with $Z_1^2 = 144.5$
at a frequency of $234.564\pm0.007$ mHz (all errors are quoted at the
1$\sigma$ confidence level). Allowing for $4.3\times10^{6}$ independent frequencies searched 
in the range 0.1--100~Hz, the probability of
chance occurrence of this signal in the absence of a real pulsation
would be $1.8\times10^{-25}$, showing that the periodicity from this source would be easily
detectable in a blind search of the X-ray photons.

The ephemeris obtained through continued radio timing of
PSR~J1819--1458 with the Parkes radio telescope at 1.4~GHz using the
TEMPO software package predicts a barycentric frequency at the center
of the observation (MJD~53830.87029) of $234.566244\pm0.000001$~mHz,
consistent with the measured X-ray frequency. Accounting
for spin-down during the observation is unimportant as the change in
frequency due to the $-3.16\times10^{-14}$~Hz~s$^{-1}$ frequency
derivative over the 12-hr observation is much less than the size of an
individual frequency bin.  The X-ray detection of
periodicity shows that the radio-derived period is indeed the true
period, and not a smaller common factor of the radio arrival times
(see McLaughlin et al. 2006).

In Figure~1, we present the 0.3--5~keV background-corrected X-ray pulse profile
formed by folding barycentered photons from the PN and both MOS
instruments using the radio ephemeris.  The same good time interval
file was used for the PN and both MOS instruments and backgrounds were
subtracted separately.  The different exposure times for the two
instruments were accounted for.  We chose to restrict this analysis to
0.3--5~keV as below 0.3~keV, the PN and MOS detectors are not well
calibrated and, above 5~keV, the background dominates. In Figure~1, we
also present the radio profile formed from adding individual bright
radio pulses with the radio ephemeris.  The X-ray pulse profile has a
0.3--5 keV background-corrected pulsed fraction of 34$\pm$6\%, defined as
$(F_{max}-F_{min})/(F_{max}+F_{min})$, where $F_{max}$ and $F_{min}$ are the
minimum and maximum values of the X-ray pulse profile. It can be well modeled as a
single sinusoid with $\chi^2_\nu \sim 1.2$ (17 d.o.f.). 
We find no significant dependence of
pulsed fraction on energy, with a pulsed fraction at energies
$0.3-1$~keV of 28$\pm7$\% and a pulsed fraction at energies from
1--5~keV of 49$\pm$10\%.

The peak of the X-ray profile, calculated by fitting a sinusoid, is at
phase $0.49\pm0.04$ and is aligned to the peak of the radio
profile at phase 0.5.  The 3~pc~cm$^{-3}$ uncertainty in the
DM of 196~pc~cm$^{-3}$ amounts to a radio arrival time uncertainty at
1.4~GHz of only 13~ms, or 0.3\% of the 4.26~s pulse period.

Our PN data can be used to place limits on the existence of X-ray
bursts from the source.  Binning the data on various timescales to
search for bursts of different widths
we find no evidence for
aperiodic variability on timescales ranging from 6~ms to the duration
of our observation.  We used photons with energies up to 10 keV,
important given the hard spectra of magnetar bursts (e.g. Gavriil et al 2004). 
Omitting frames with high particle background, we 
find that none of the $4\times10^{6}$ frames contains more than two
counts within the 20$^{\prime\prime}$ region, with three frames
containing exactly two counts. These numbers are entirely consistent
with Poisson statistics. The number of events in a single frame which
would deviate from a steady flux at the 3-$\sigma$ level is 3
photons. We therefore adopt this as the upper limit on the X-ray flux
of any burst of width 6~ms or less. Assuming a spectrum of the same
shape as that fitted for the overall source (column 2 of Table~1; see Section~2.2)
limits the observed fluence of any burst to $\la
8\times10^{-12}$~ergs~cm$^{-2}$ (0.3--5~keV), corresponding to an
absorbed flux limit of $\la 3\times10^{-9}$~ergs~cm$^{-2}$~s$^{-1}$
(0.3--5~keV) if we assume that the X-ray burst would last for 3~ms,
the width of the radio pulse. This is a factor of five better than the
limits placed by Reynolds et al. (2006).  At a distance of 3.6~kpc,
these limits correspond to a total energy of $2\times10^{34}$~ergs and
luminosity of $1\times10^{37}$~ergs~s$^{-1}$. Therefore, any bursts,
at least during the 12-hour duration of our observation, must contain
much less energy than typical X-ray magnetar bursts (e.g. Woods \& Thompson
2006) if their spectra is indeed similar to that
described in column 2 of Table~1.

\subsection{Spectral Analysis}
\label{spectrum}

For the spectral analysis, we used PN photons with PATTERN $\le$ 4 (i.e. single and double events) and
MOS1 and MOS2 photons with PATTERN $\le$ 12. Source and background
spectra were extracted from the same regions used for the timing
analysis and the spectral response matrices were created with the SAS
{\tt mkrmf} and {\tt mkarf} tools, using the bad-pixel file built for
our observation.  The PN and MOS spectra were only used in the
0.4--2\,keV energy range, a smaller range than that used for the timing analysis due to the
greater dependence of spectral fitting on background
spectra.
In the spectral analysis we used both rebinned and
unbinned spectra. Spectra were rebinned for MOS and PN
by a factor of two so that the energy resolution was not
oversampled by more than a factor of three (using the specific
response matrix built for this observation). Furthermore, we rebinned
in order to have at least 30 counts per bin so that we could
use the
$\chi^2$ statistic\footnote{http://heasarc.nasa.gov/xanadu/xspec/manual/manual.html}.

We first modeled the PN spectrum. We tried several different models 
and found that a single component fit was not
possible. Fitting with a single absorbed\footnote{If not otherwise
specified, abundances were assumed to be solar and fixed at the
values in Lodders (2003).} blackbody (as in Reynolds et al.~2006),
we found $N_H\sim 4.0\times 10^{21}$\,cm$^{-2}$ and $kT\sim0.14$\,keV
with a $\chi^2_\nu > 2.0$ (see Figure~2), while for an absorbed
powerlaw, we obtained $N_H\sim 1.3\times 10^{22}$\,cm$^{-2}$ and
$\Gamma\sim8.2$ with a $\chi^2_\nu >$2.1 (both fits have 53
d.o.f). 
 These values are consistent with the Reynolds et al. (2006)
best-fit parameters (0.5--8~keV) of $N_H = 7^{+7}_{-4} \times 10^{21}$
cm$^{-2}$ and $kT=0.12 \pm 0.04$\,keV (for the absorbed blackbody
model) and $\Gamma\sim9.5$ (for the power-law model).  An inspection
of the residuals revealed that our high $\chi^2_\nu$ values were due
to the presence of strong spectral features around 1\,keV and a
weaker one around 0.5\,keV. 

We carefully checked whether these features might be due to
calibration issues, to our source and background extraction regions or
to residual particle flares and/or particles hitting the detector. 
We reliably excluded all of these issues by studying in detail the
{\it XMM-Newton} calibration lines (see Figure~3), extracting
the source and background photons from several different
regions and investigating the spectrum using only PATTERN =
0 counts (i.e. only isolated events).
However, we note that the 0.5\,keV feature is very
close to the Oxygen edge energy, and an overabundance of Oxygen in
the direction of the source
could be responsible.
To further investigate this
possibility, we fit our spectrum with an absorbed blackbody model using
three different photoelectric cross-sections, those of
Ba{\l}uci\'nska-Church \& McCammon (1992, 1998) and of Verner
et. al. (1996). This was aimed at studying the dependence of the
significance of our 0.5\,keV line on the chosen photoelectric
cross-section, which drives the shape of the edge.  We found that the
residuals around 0.5\,keV remain in all cases, although less
significantly if we use the Verner et. al. (1996) cross-section. For
that reason, we assumed this cross-section for all our spectral
fitting, and we omitted all photons in the 0.50--0.53~keV energy range (two
bins in our rebinned PN spectrum) from our modeling. We tentatively
conclude that this line is due to the Oxygen edge, but only future
deeper observations will unambiguously confirm this. 

\begin{table*}
\begin{center}
\caption{Spectral fits for PSR~J1819--1458 with EPIC-PN}
\vspace{0.3cm}

\begin{tabular}{lclclclc}
\hline
\hline
  \multicolumn{2}{c}{Blackbody (BB) plus Neon} & \multicolumn{2}{c}{BB plus Gaussian}
& \multicolumn{2}{c}{BB plus edge} & \multicolumn{2}{c}{BB plus cyclotron}\\
\hline
& & & & & & & \\
$N_{H}$  & 0.59$^{+0.06}_{-0.04}$ &         &   0.75$^{+0.12}_{-0.09}$  &
& 0
.57$\pm0.06$   &    & 0.81$^{+0.09}_{-0.08}$  \\
$N_{Ne}$  &  6$\pm1$ &    E$_{G}$        &
1.11$^{+0.04}_{-0.03}$
 & E$_{e}$ &  0.92$^{+0.03}_{-0.01}$  & E$_{cy}$  & 0.99$^{+0.03}_{-0.02}$
\\
\\
    &          & $\sigma_{G}$    &     0.21$^{+0.03}_{-0.06}$      &   &
& w$_{c
y}$ &  0.37$^{+0.03}_{-0.06}$  \\
    &            &   $\tau_{G}$   &   150$\pm{60}$  &$\tau_{e}$ &
0.67$\pm0.14$
  & d$_{cy}$ & 1.2$\pm0.2$ \\

& & & & & & & \\
$kT$ & 0.144$^{+0.008}_{-0.006}$ & &  0.136$^{+0.012}_{-0.008}$   &    &
0.150$^{+0.
005}_{-0.006}$ & &  0.144$^{+0.008}_{-0.006}$ \\
Abs. Flux&  1.5$^{+0.3}_{-0.5}$ & & 1.5$^{+0.6}_{-0.8}$   & &
1.5$^{+0.3}_{-0.6}$  & &
 1.5$^{+0.3}_{-0.4}$  \\
Unab. Flux & 15$\pm2$      & &  26$^{+2}_{-8}$ & &
12$^{+2}_{-2}$    & & 34$^{+4}_{-9}$    \\
$\chi^2_\nu$ (d.o.f.)   & 1.20 (50)  & & 1.19 (48)   & & 1.17 (49)  & &
1.13 (48)  \\

\hline
\end{tabular}
\tablecomments{Fluxes are calculated in the 0.3--5\,keV energy range,
  and reported in units of
  $10^{-13}$~ergs\,s$^{-1}$\,cm$^{-2}$. $N_{H}$ is in units of
  $10^{22}$cm$^{-2}$ and $N_{Ne}$ is in solar units (always assuming
  solar abundances from Lodders 2003). The photoelectric cross-section
  of Verner et al. (1998) has been used for all fits. The values of $kT$
  (blackbody temperature), E$_{G}$ (Gaussian line energy),
  $\sigma_{G}$ (Gaussian line width), $E_e$ (edge threshold energy),
  E$_{cy}$ (cyclotron line energy) and w$_{cy}$ (cyclotron line width)
  are in units of keV. The Gaussian line depth $\tau_G$, edge depth
  $\tau_e$ and fundamental cyclotron line depth $d_{cy}$ are
  dimensionless. 
  Errors are at the 1$\sigma$
  confidence level. XSPEC models used are (from left to right): {\tt
    vphabs*bbody, phabs*gabs*bbody, phabs*edge*bbody} and {\tt
    phabs*cyclabs*bbody}.}
\end{center}
\label{spec}
\end{table*}

Keeping the number of components as low as possible, we tried to model
our rebinned PN spectrum with either an absorbed blackbody (XSPEC model {\tt bbody}) or
power-law ({\tt powerlaw}), in addition to modeling the $\sim$1\,keV feature in several
ways: leaving free the abundances of the most abundant elements in the
interstellar medium (ISM) with lines around 1\,keV (e.g. Ne, N, Mg,
Fe), adding a Gaussian function (XSPEC model {\tt gabs}), a Lorentzian
function ({\tt lorentz}), an absorption edge ({\tt edge}) or a cyclotron
resonant scattering model ({\tt cyclabs}).
Among our trials, only the models reported in
Table\,1 gave satisfactory values of $\chi^2_\nu\sim1$. We show the
blackbody plus Gaussian fit and residuals in Figure~4. Note that there
is evidence in the residuals for narrow lines within our broad 1 keV
line, implying that the line may be a blending of narrower
lines. However, the addition of several single narrow lines is not
statistically significant, and future data with better statistics are
necessary for investigating this issue.

We ran Monte Carlo simulations of 2$\times10^4$ spectra (see Rea et
al.~(2005, 2007) for details) to estimate the significance
of the spectral feature (as suggested by Protassov et
al.~2002)\footnote{Even though the F-test is still widely used to
assess the significance of features in X-ray spectra, it has been
shown that this is not statistically correct (Protassov et
al.~2002).}.  The spectra were simulated using models corresponding to columns 2 and 4 from
Table 1 with the absorbed blackbody parameters varying within their
3$\sigma$ errors (see Table\,1), absorption line parameters
completely free to vary, and fixing the number of counts of each
spectra at the PN count value for PSR\,J1819--1458.  After having generated
10$^4$ spectra for each of the two models, we counted how many of
these simulated spectra showed an absorption feature at any energy,
width or depth, only due to statistical fluctuations. None of the
simulated spectra presented a Gaussian absorption line at any energy
or width with a depth $\tau_G > 30$, and no
cyclotron component was detected with a depth $>0.3$ for any energy or
width.  We therefore infer the significance of the line at 1\,keV to
be $> 1/10^4$, corresponding to a $>$ 99.99\% probability (i.e. at least 4$\sigma$) that this
line is not due to statistical fluctuations.

To check whether the line was also present in the MOS data, we
simultaneously fit the PN and both MOS1 and MOS2 rebinned spectra (see
Figure~5).  All models in Table~1 fit the PN, MOS1 and MOS2
data well. However, because of the lower number of MOS counts, our 1~keV
line was not significant in the MOS spectra, especially because the
rebinning we did in order to use the $\chi^2$ statistic left only a
handful of bins within the line. As a further check, we then used the
unbinned PN and MOS spectra and the C-statistic (Cash 1979)  to assess the
goodness of an absorbed blackbody fit, without the inclusion of any
line. We found that for $N_H=0.5\times10^{22}$\,cm$^{-2}$ and
$kT=0.14$\,keV, the C-statistic is 1009.776 using 969 PHA bins. A
Monte Carlo simulation of the fit showed that the probability of
having a C-statistic $<$ 1010 is 98.0\%. Note that only if the C-statistic 
probability is smaller then 50\% can the model be accepted. This
shows that an absorbed blackbody model alone can not reproduce the
data, and hence that a line is also present in the MOS spectra.  We show 
in Figure\,5 the PN, MOS1 and MOS2 spectra modeled with an absorbed
blackbody plus a Gaussian (column 2 in Table\,1), with a more severe PN
rebinning (with at least 40 counts per bin), to show
the agreement between the three instruments.  Note that all the models in
Table~1 satisfactorily fit the previous {\it Chandra} data (Reynolds
et al. 2006), although the absorption line is not statistically
significant in those data because of the reduced number of counts.

Furthermore, we checked whether the spectral feature parameters and
significance were dependent on the spectral binning or on the
continuum model.  We found that the spectral feature does not vary
significantly when we vary the amount of spectral binning (see also Figure\,5), but
that it is rather dependent on the chosen continuum (as is generally
the case for every spectral feature). In fact, if instead of a blackbody
 model, we assume a power--law continuum, 
the feature is still detectable with similar
spectral properties but at a lower significance of 98.9\% (roughly 2.5$\sigma$).

Hence, if we assume a blackbody continuum model
(reasonable for an isolated neutron star) we can show with high
confidence that a spectral feature is present in our data.
However, our statistics do not allow us to statistically prefer one
model to the others and reliably ascertain its nature.
Furthermore, if future, more sensitive observations reveal
a different source continuum, the significance of the
feature may need to be revised.

We also performed phase-resolved spectroscopy over two intervals of
0.5 in phase centered on the peak and the minimum of the pulse
profile, respectively.  We did not find any significant spectral
variability between these two phase intervals at the 2$\sigma$ level.
However the low number
of counts in each spectrum does not allow us to reach any reliable
conclusions about the pulse-phase dependence of the spectral
parameters.

\section{Discussion}

\subsection{Interpretation of the spectral feature}

If we accept that the spectral feature is significant, the two
main possible interpretations are as an atomic line or as a cyclotron
resonant scattering line. 

If the line is atomic, then it could be due either to the neutron star
atmosphere or, but less probably, to a peculiar abundances in the ISM
in that direction (or perhaps even a combination of the two). At
1\,keV, it might then be due to the N edge (although improbable given
its low abundance), Ne edge (around 0.9~keV), Ne XI K-edge (at
1.19\,keV), or some Fe-L lines, including Fe XX,
Fe XXI and Fe XXII or Fe XXIII (all around 0.9\,keV). Note that Fe lines are common in other
kinds of neutron stars, though mostly coming from accretion disks in
binary systems rather than from the neutron stars themselves (see
e.g. Cottam et al. 2002). The structure
present in the residuals (see Figure~4 middle panel), which in our data
is not significant, might be due to
a blending of narrow lines which we could only model with a broad
line due to our limited number of counts. 

If the feature is
due to proton cyclotron resonant scattering, the magnetic field inferred would be
$B_{cy}=1.6E_{cy}({\rm keV})/y_G~10^{14}$\,G, where $y_G =
(1-2GM/c^2R)^{1/2}$ is the gravitational redshift factor ($M$ and $R$
are the neutron star mass and radius, respectively). Assuming
canonical values for $M$ and $R$ of 1.4~$M_\odot$ and 10~km, we find
$B_{cy} = 2\times10^{14}$~G.  This is slightly higher than the dipolar
surface magnetic field inferred through radio timing through the standard formula $B=3.2\times10^{19}{P\dot{P}^{1/2}}=5\times10^{13}$~G.  However,
this expression can be considered an order-of-magnitude estimate as it
assumes a purely dipolar field, a neutron star radius of 10~km, a
moment of inertia of $10^{45}$~g~cm$^{2}$ and an angle between the
spin and magnetic axes $\alpha = 90$~degrees. 
 Accounting for
this and assuming $\alpha = 30$~degrees would make the timing- and
cyclotron-inferred fields consistent.  
 In addition, the width and
depth of the line are consistent with the predictions of Zane et
al. (2001) for proton-cyclotron absorption in highly magnetized
neutron stars.  We cannot detect any harmonics to the fundamental
cyclotron line because our spectrum is
background dominated above 2\,keV.
However, as observed in some accreting
sources (Heindl et al. 2004) and investigated for isolated neutron stars
by Liu et al. (2006), in some cases the first harmonic is
as deep as the fundamental cyclotron frequency. Therefore, it is possible, although
unlikely, that 
the 1\,keV feature is the first
harmonic, with the 0.5~keV fundamental
coincident with the depression in the spectrum that we have
interpreted as due to an overabundance of Oxygen.
We note that the observed lines are unlikely to be
 due to electron-cyclotron absorption as the
inferred magnetic field would be a factor of 2000 lower
(i.e. the ratio of the proton to electron mass) 
and incompatible with that measured through radio timing.

\subsection{Relationship to other classes of neutron stars}

Our detection of periodicity at the radio period shows that the X-ray
source reported by Reynolds et al. (2006) is undoubtedly the
counterpart to PSR~J1819--1458. The pulsed fraction and sinusoidal
pulse shape are similar to what is observed for other middle-aged
X-ray-detected radio pulsars such as B0656+14 (e.g. De Luca et
al. 2005), which has been observed to have similar radio
properties to the RRATs (Weltevrede et al.~2006). 

The thermal emission from PSR~J1819--1458 is consistent with a cooling
neutron star. However, the temperature from our blackbody fit appears
slightly higher than temperatures derived from blackbody fits for
other neutron stars of similar ages (see discussion by Reynolds et
al.~2006).
Note that it is possible that PSR~J1819--1458 was born
spinning at a sizable fraction of its present period of 4.26~s.  In
this case, as discussed by Reynolds et al. (2006), the characteristic age could be a
considerable overestimate and the inferred temperature could be
completely consistent with its age.  Note that characteristic ages
have been shown to be misleading for several other pulsars
(e.g. Gaensler \& Frail 2000; Kramer et al.~2003).

Including PSR~J1819--1458, eight high-magnetic field radio pulsars
(i.e. $B > 1\times10^{13}$~G)
have now been observed at X-ray energies. 
Two, J1846--0258 and B1509--58, are bright non-thermal sources (Mereghetti et al. 2002; Cusumano et al. 2001),
as expected given their young ages (less than 2000
yr). PSR~J1119--6127 is a bright thermal X-ray emitter with
unusual properties including a large pulsed fraction and narrow
pulse (Gonzalez et al. 2005).
PSR~J1718--3718, with magnetic field of $7\times10^{13}$~G,
has been detected at X-ray energies, but the faintness of the
counterpart does not allow detailed spectral modeling or a
constraining limit on pulsed fraction (Kaspi \& McLaughlin 2005). No
X-ray emission has been detected from PSRs J1814--1744, B0154+61 or
J1847--0130, which has the highest inferred surface dipole magnetic
field ($9\times10^{13}$~G) measured to date for any radio pulsar
(Pivovaroff et al. 2000; Gonzalez et al. 2004; McLaughlin et al. 2003). Radio pulsar X-ray emission
properties clearly vary widely, even for objects with very similar spin-down
properties.

While the spectrum and luminosity of PSR~J1819--1458 
argue against a relationship with
magnetars, it is possible that
PSR~J1819--1458 could be a transition object between
the pulsar and magnetar source classes.  The soft X-ray spectrum does
have a comparable temperature to the quiescent state of XTE~J1810--197
($kT \sim 0.15-0.18$ keV; Ibrahim et al.~2004; Gotthelf et al.~2004).
However, the radio emission characteristics of these two neutron stars
are quite different.

While resonant cyclotron features are regularly observed from X-ray binary
systems (e.g. Truemper et al. 1978; Nakajima et al. 2006), the
detection of such features from isolated neutron stars is quite
unusual. Sanwal et al. (2002) \& Mereghetti et al. (2002)
 discovered  harmonically spaced
absorption lines from 1E~1207.4--5209, a radio-quiet X-ray pulsar with
a 424-ms spin period and timing-derived characteristic age and
inferred surface dipole magnetic field strength of $3\times10^{5}$~yr
and $3\times10^{12}$~G, respectively. Analyzing a deeper observation of the source,
Bignami et al. (2003) attributed these lines to
electron-cyclotron absorption. However, 
the significance of two of the lines present in its spectrum has been strongly
questioned (Mori et al.~2005). 

Broad absorption lines, similar to those seen for
PSR\,J1819--1458, have been observed
for six out of seven XDINS 
(see reviews by van Kerkwijk \& Kaplan 2007 and Haberl 2007). For most of these neutron stars, the
lines can be interpreted as due to neutral hydrogen transitions in
highly magnetized atmospheres.  Ho et al. (2003) and Van Kerkwijk \& Kaplan (2007) argue that
the transition energy is similar to the proton
cyclotron energy for magnetic fields of the order of PSR~J1819--1458's.
The X--ray
spectrum of PSR\,J1819--1458 is very similar (although with a slightly
hotter blackbody temperature) to the XDINSs, although so far no
convincing evidence for radio bursts has been detected for any of
those thermally emitting neutron stars (Kondratiev et al., in
preparation).

One outstanding question is 
why absorption lines of this kind, whether due to the atmosphere or
to cyclotron resonant scattering, have been observed from only a
handful of X-ray emitting isolated neutron stars. As suggested by Mereghetti et al.
(2002), the age of the
neutron star could be one key factor. Young objects are dominated by
non-thermal emission, but older ones may be too faint for X-rays to be
detectable, making X-ray bright, middle-aged pulsars the best
candidates (as is the case of the XDINSs and of
PSR\,1819--1458). Note, however, that no such absorption lines have
been found for the X-ray bright, middle-aged PSR\,B0656+14, despite
deep searches both with {\it Chandra} (Marshall \& Schultz 2002) and
{\it XMM-Newton} (De Luca et al. 2005) and the theoretical predictions
for cyclotron and/or atmospheric features expected in its
emission. The explanation could well depend on the
viewing geometry.

More sensitive observations are necessary to confirm the presence of
absorption features in the spectrum of PSR~J1819--1458. Longer X--ray
observations are needed to understand whether the broad 1~keV line could be a
blending of narrower
features. This would then argue for an atomic origin for the
line. Longer observations will also allow  
phase-resolved
spectroscopy. Simulations show that a 100-ks observation would allow us to 
achieve both of these goals.
 Because the strength of the cyclotron line depends on
the angle between the observer and the magnetic field, we expect phase variation of
cyclotron features.
If the feature we detect is indeed due to
proton-cyclotron absorption, it provides an invaluable means of testing the assumptions
implicit in the characteristic magnetic fields derived through radio timing
measurements and a unique independent measurement of the magnetic field of an isolated neutron star.

\section{Conclusions}

We have discovered X-ray pulsations at the 4.26-s period inferred from
radio timing of PSR~J1819--1458. 
The properties of these pulsations are similar to those
observed for other middle-aged radio pulsars detected at X-ray
energies. While the RRATs are characterized by sporadic radio emission,
 we do not detect any X-ray bursts or aperiodic variations
throughout the observation and can place the most stringent limit to
date of $\le 3\times10^{-9}$~ergs~cm$^{-2}$~s$^{-1}$ on the absorbed
0.3--5~keV flux of any X-ray bursts.  We have characterized the
spectrum of this source and find that it is well-described by an
absorbed blackbody with $kT = 0.14$~keV in addition to an absorption
line around 1\,keV, with total absorbed flux of
$1.5\times10^{-13}$~ergs~cm$^{-2}$~s$^{-1}$ (0.3--5~keV).  We note,
however, that the presence of this absorption feature is highly
dependent on the choice of continuum model and needs further X-ray
observations to be confirmed. This object is the only RRAT so far to be
detected at X-ray energies. X-ray observations of the other objects in this
source class are essential for a complete picture of how they relate to other
neutron star populations.

\acknowledgments

MAM thanks the Parkes Multibeam Survey team for radio observing assistance and
 Pete Woods for helpful discussions on XMM data analysis.
This work was supported by NASA through {\it XMM-Newton} Observer Program
grant NNX06AG20G. N.R. is supported by an NWO Postdoctoral Fellowship
and a Short Term Visiting Fellowship awarded by the University of
Sydney, and kindly thanks M. M\'endez for useful suggestions.

\begin{figure}
\epsscale{.60}
\plotone{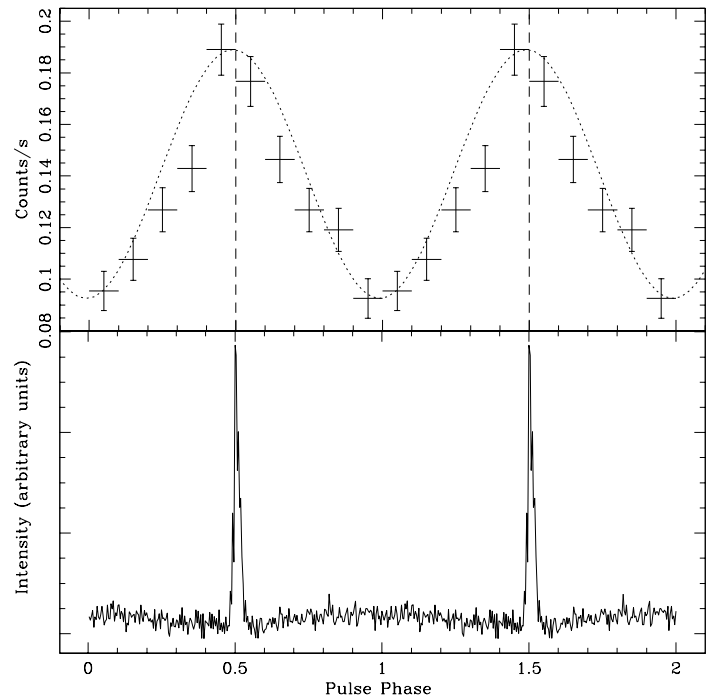}
\caption{Top: X-ray
pulse profile (0.3--5 keV) for 23-ks of EPIC-PN data and 37-ks of MOS1 and MOS2
data
 on PSR~J1819--1458. The dotted line shows the best-fit sinusoid and the vertical dashed line indicates the phase of the peak of the radio pulse.
Bottom: radio profile formed from 114 pulses detected in 6 hr of observation
at 1.4~GHz with the Parkes telescope in Australia. The non-uniform baseline is likely due to
instrumental digitization effects. Both profiles have been folded
using the radio ephemeris. In both plots, two cycles of pulse phase are shown.
\label{profile}}
\end{figure}

\begin{figure}
\epsscale{0.80}
\plotone{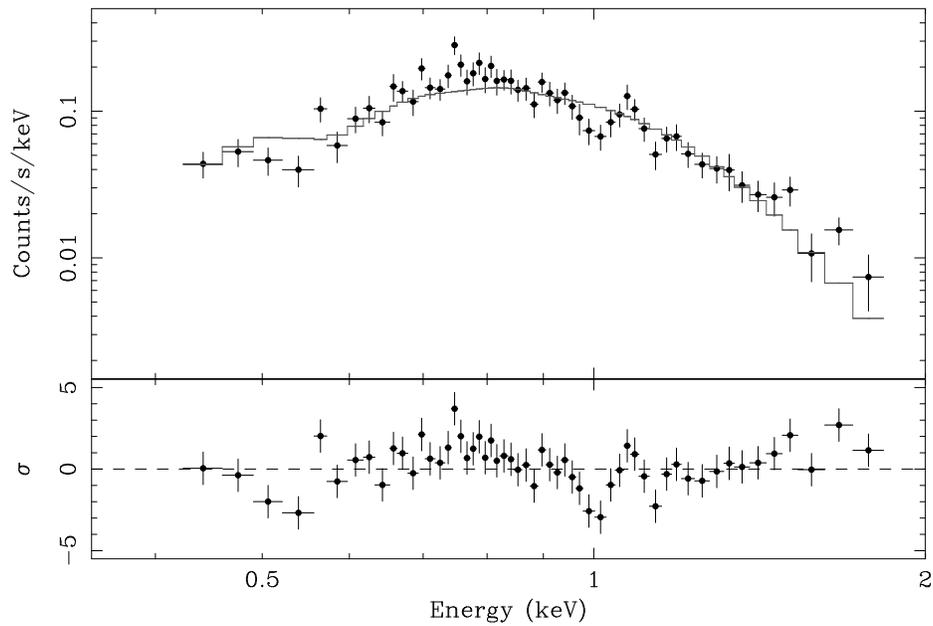}
\caption{
    Top: {\it XMM-Newton} EPIC-PN spectrum of PSR~J1819--1458, modeled as an
    absorbed blackbody. The points 
 indicate the data, while the solid line shows the corresponding
best-fit model.
    Bottom: residuals of the 
    model.} 
\end{figure}

\begin{figure}
\epsscale{0.80}
\plotone{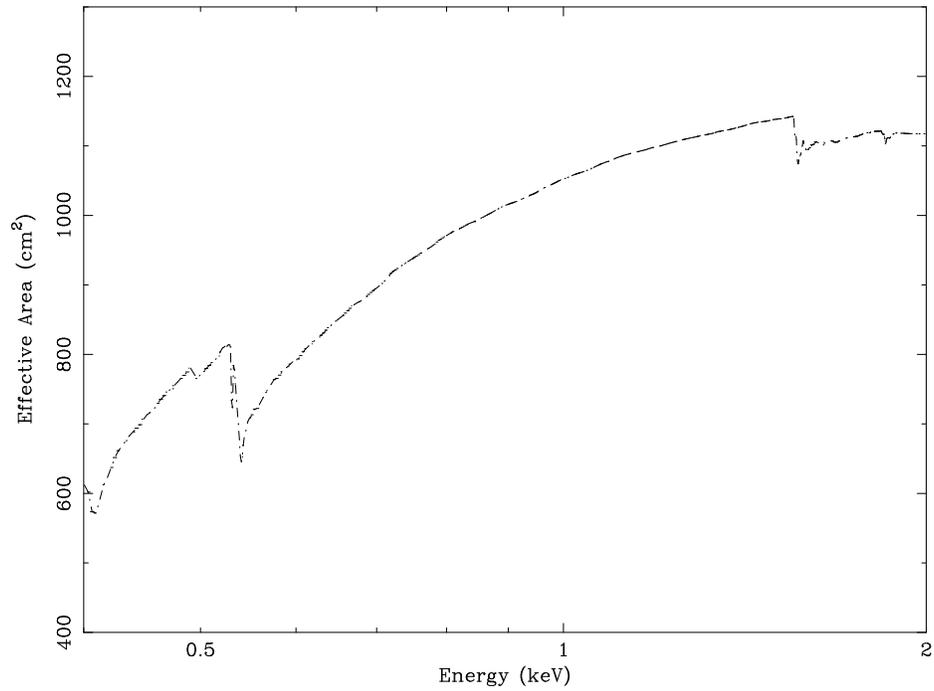}
\caption{The effective area, or total response efficiency, for EPIC-PN
with the medium filter versus incident photon energy. Note that there is no instrumental feature in
the effective area curve at
1~keV. The calibration feature near 0.5~keV is too weak to explain the feature in our spectrum
close to this energy (see Haberl et al. 2006 for more details on {\it XMM-Newton} calibration.)}
\end{figure}

\begin{figure}
\epsscale{0.80}
\plotone{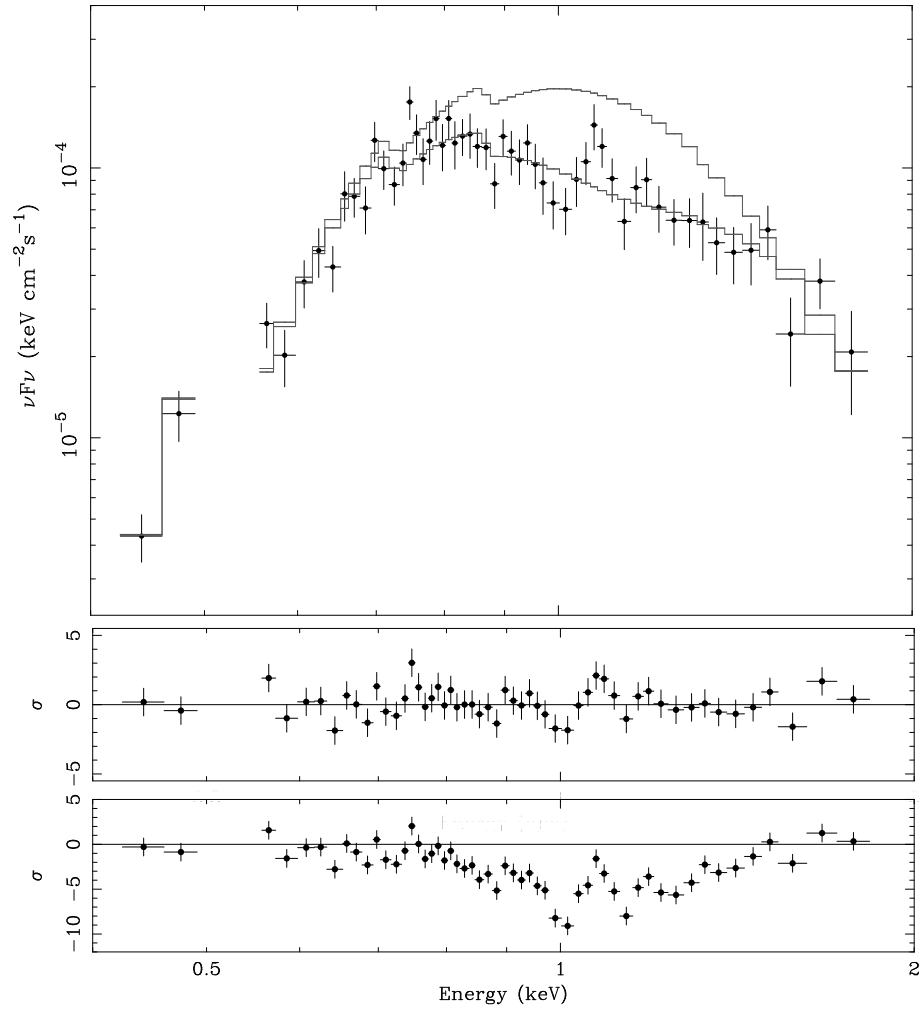}
\caption{
    Top: {\it XMM-Newton} EPIC-PN $\nu F_\nu$ spectrum of PSR~J1819--1458, modeled as an
    absorbed blackbody and Gaussian line (column~2 of Table~1). The points indicate the data, while the solid line shows the corresponding
best fit model.  Top line shows the absorbed
blackbody model component of the fit alone. We have omitted all photons with energies from
0.5--0.53~keV as we believe that the feature in this range is due to the Oxygen edge (see Section~2.2).
    Middle:
    residuals of the absorbed blackbody and Gaussian line model. Bottom: residuals of the model without
the inclusion of the Gaussian line.}
\end{figure}

\begin{figure}
\epsscale{0.80}
\plotone{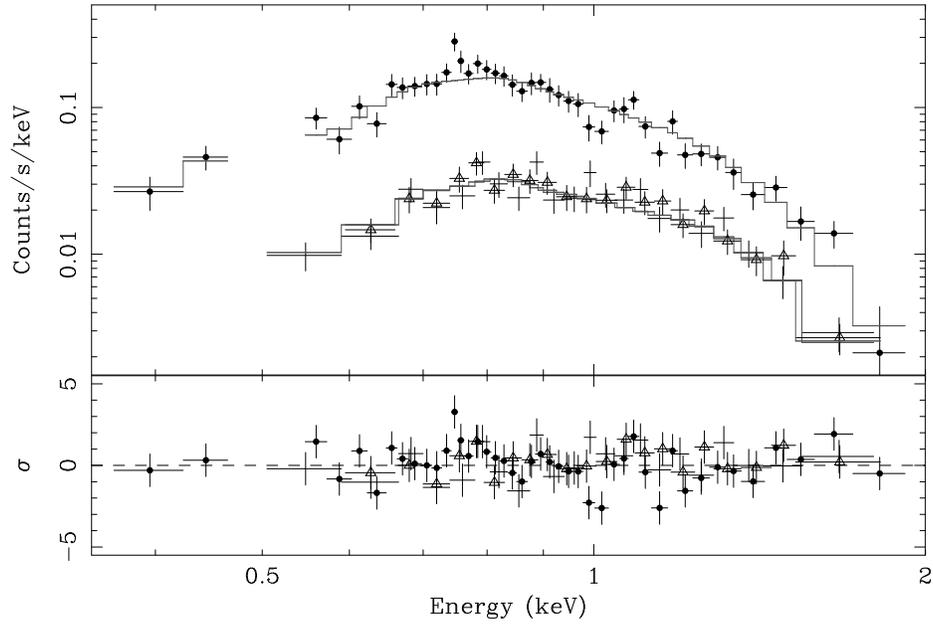}
\caption{Top: {\it XMM-Newton} EPIC-PN (filled circles), MOS1 (black
  cross), and MOS2 (triangles) spectra  for
  PSR~J1819--1458, modeled as an absorbed blackbody and Gaussian line
  (column~2 of Table~1). Grey lines show the fitted model. The PN data
have been binned with 40 counts per bin and the MOS1 and MOS2 data with 30 counts per bin to show that the best-fit model is not dependent on the spectral binning parameters. We have omitted all photons with energies from
0.5--0.53~keV as we believe that the feature in this range is due to the Oxygen edge
 (see Section~2.2). Bottom: residuals of 
the above fits.}
\end{figure}

\end{document}